\documentstyle[preprint,aps]{revtex}
\begin{document}
\draft
\title{Reaction rates for Neutron Capture Reactions to
C-, N- and O-isotopes to the neutron rich side of stability}
\author{H.~Herndl, R.~Hofinger, J.~Jank and H.~Oberhummer}
\address{
	Institut f\"ur Kernphysik, Technische Universit\"at Wien,
	Wiedner Hauptstra{\ss}e 8-10, A-1040 Wien, Austria}
\author{J.~G\"orres, M.~Wiescher}
\address{
	Department of Physics, University of Notre Dame, Notre Dame,
	Indiana 46556, USA}
\author{F.-K.~Thielemann}
\address{Department f\"ur Physik und Astronomie, Universit\"at Basel,
Klingelbergstr. 82, CH-4056 Basel, Switzerland}
\author{B.A.~Brown}
\address{
	Department of Physics and Astronomy, Michigan State University,
    East Lansing, USA}
\date{\today}
\maketitle
\begin{abstract}
The reaction rates of neutron capture reactions on light nuclei are important
for reliably simulating nucleosynthesis in a variety of stellar scenarios.
Neutron capture reaction rates on neutron-rich C-, N-, and O-isotopes
are calculated in the framework of a hybrid compound and direct capture model. 
The results are tabulated and compared with the results of previous
calculations as well as with experimental results. 
\end{abstract}
%
%
%
%
%
\section{Introduction}
Neutron capture processes on neutron-rich C-, N- and O-isotopes
play an important role in astrophysical scenarios ranging from 
nucleosynthesis in the stellar helium and carbon burning stages
to possibly inhomogeneous Big Bang models. To simulate the nucleosynthesis of light
isotopes between carbon and neon a detailed understanding of the these
neutron capture reactions is therefore essential. This paper intends to
derive a consistent set of stellar neutron capture reaction rates for
neutron-rich carbon, nitrogen and oxygen isotopes, based on the latest
experimental and theoretical information.

In stellar helium core burning in massive Red Giant stars neutrons are
abundantly produced via the $^{14}$N($\alpha$,$\gamma$)$^{18}$F($\beta^+\nu$)
$^{18}$O($\alpha$,$\gamma$)$^{22}$Ne($\alpha$,n) reaction
sequence~\cite{Ibe82,KWG94}. In the early helium core evolution $^{14}$N and
$^{18}$O are initially depleted, while initiating the neutron production.
The neutrons subsequently trigger the weak s-process component
which leads to the production
of intermediate mass nuclei around A=100. Neutrons, however, can also be
captured on $^{12}$C and $^{16}$O which are abundantly produced in helium
burning.

In the very last phase of helium core burning, the core expands and
outer layers with high $^{14}$N and $^{18}$O abundances cause an increase
in neutron production. Neutron capture on these isotopes
however may act as neutron poison and may also change the light isotope
abundances. This depends critically on the reaction cross section in the
energy range between 25 and 200\,keV.

He-shell burning in low mass asymptotic giant branch (AGB) stars
has been proposed as the site for the main component of the s-process
\cite{HoI88,KGB90,SGB95}. Neutron production is triggered in the He-burning
shell by the $^{13}$C($\alpha$,n) reaction on $^{13}$C being ingested
by convective processes during the thermal pulses. It has been suggested that
neutron induced processes on the initial and the additionally ingested carbon,
nitrogen, and oxygen abundances may have considerable influence on light
isotope nucleosynthesis in the thermal pulse~\cite{FGJ92,JSL92}. Recent
stellar model calculations \cite{SGB95} indicate that the
reaction cross sections for the neutron capture processes need to be known
in the energy range between 5\,keV and 30\,keV.

In the framework of inhomogeneous Big Bang models (IMs)
high neutron flux induces primordial nucleosynthesis which bridges the mass 5 and
mass 8 gap~\cite{app88}. Subsequent neutron capture
processes on neutron-rich carbon, nitrogen, and oxygen isotopes may bypass
the long-lived $^{14}$C and trigger a primordial r-process~\cite{rau94,TSO93,TSO94,OKB97}.
The efficiency for the production of heavy elements in such a scenario
depends sensitively on the respective neutron capture rates for these
light isotopes. Therefore the neutron capture cross sections have to be
determined over a wide energy range up to 1.0\,MeV.

Over the last few years considerable effort has been made to determine
the neutron capture reaction rates for the C-, N-, and O-isotopes
experimentally as well as theoretically. With the present paper we
attempt to summarize the
experimentally determined neutron capture rates. We present new calculations
in the framework of a hybrid compound
nucleus and direct capture model and compare the results with the
experimental data as well as with previous calculations using the
direct capture model \cite{rau94} and
the statistical Hauser-Feshbach
model \cite{CTT91,RTK97}. The same models were used to
determine in addition neutron capture reaction rates on $\beta^-$-unstable
neutron-rich C-, N-, and O-isotopes.

In the following Sect.~2 we will present the formalism used in
calculating the reaction cross sections and reaction rates. We
also discuss the experimental and theoretical input parameters for
the calculations of the different cross sections and reaction rates.
In Sect.~3
the results of our calculations are compared with previous predictions
both with a direct capture and a Hauser-Feshbach model. If available we
also compare our results with experimental data.
Finally, in the last section the results are summarized and discussed.
\section{Calculation of the Reaction Rates}
The cross section for neutron capture processes is dominated by
the non-resonant direct capture (DC) process and by contributions from
single resonances which correspond to neutron unbound states in the
compound nucleus (CN). For calculating the different reaction contributions
we used a simple hybrid model: the non-resonant contributions were
determined by using a direct capture model, the resonant contributions
were based on determining the resonant Breit-Wigner cross section. In the
case of broad resonances interference terms have to be taken into account.
To determine the neutron capture cross sections on the $\beta$-unstable
nuclei the necessary input parameter for the calculations
(masses, Q-values, spin-parity assignments of bound states
and resonances, excitation energies, spectroscopic factors, density
distributions, scattering data) were taken from experimental data.
When no experimental data were available we used theoretical
values, mainly derived from the shell model (see below).

The total reaction rate is given by
\begin{equation}
N_{\rm A} <\sigma v>_{\rm tot} =
N_{\rm A} <\sigma v>_{\rm r} +
N_{\rm A} <\sigma v>_{\rm nr} +
N_{\rm A} <\sigma v>_{\rm int} \quad ,
\end{equation}
where the three terms represent the resonant contribution
$N_{\rm A} <\sigma v>_{\rm r}$, the non-resonant contribution
$N_{\rm A} <\sigma v>_{\rm nr}$ and the interferences
$N_{\rm A} <\sigma v>_{\rm int}$. Each contribution will be
explained in the following sections.

Another important quantity is the Maxwellian averaged cross
section.
For a temperature $kT$ it is defined by   
\begin{equation}
\frac{<\sigma v>_{kT}}{v_T} = \frac{2}{\sqrt{(kT)^2}}
\int_0^{\infty} E \sigma_{{\rm n},\gamma} (E) \exp{\left(-\frac{E}{kT}\right)} dE
\quad .
\end{equation}
\subsection{Resonant Reaction Contributions}
The cross section of a single isolated resonance in neutron capture
processes is well described by
the Breit-Wigner formula~\cite{Bre36,Bla62}
\begin{equation}
\label{BW}
\sigma_{\rm r}(E) = 
\frac{\pi \hbar^2}{2 \mu E}
\frac{\left(2J+1\right)}{2\left(2j_{\rm t}+1\right)} 
\frac{\Gamma_{\rm n} \Gamma_{\rm\gamma}}
{\left(E_{\rm r} - E\right)^2 + 
\left(\frac{\Gamma_{\rm tot}}{2}\right)^2} \quad ,
\end{equation}
where $J$ and $j_{\rm t}$ are the spins of the resonance level and the
target nucleus, respectively, $E_{\rm r}$ is the resonance energy.
The partial widths of the entrance and exit channel
are $\Gamma_{\rm n}$  and $\Gamma_{\rm\gamma}$, respectively.
The total width $\Gamma_{\rm tot}$
is the sum over the partial widths of all channels.
The neutron partial width $\Gamma_{\rm n}$ can be expressed in terms of the
single-particle spectroscopic factor $S$ and the single-particle width
$\Gamma_{\rm s.p.}$ of the resonance state~\cite{wie82,her95}
\begin{equation}
\label{SF}
\Gamma_{\rm n} = C^2 S \times \Gamma_{\rm s.p.} \quad,
\end{equation}
where $C$ is the isospin Clebsch-Gordan coefficient.
The single-particle width $\Gamma_{\rm s.p.}$ can be calculated
from the scattering phase shifts of a scattering potential with the
potential depth being determined by matching the resonance energy.

The gamma partial widths $\Gamma_{\rm\gamma}$ are calculated from
the electromagnetic
reduced transition probabilities B($J_{i}\rightarrow J_{f}$;L) which
carry the nuclear structure information of the resonance states
and the final bound states \cite{bru77}.
The reduced transition rates were computed within the framework of the
shell model.

Most of the transitions in this work are M1 or E2 transitions.
For these the relations are
\begin{equation}
\Gamma_{\rm E2} [\rm{eV}] = 8.13 \cdot 10^{-7} E_{\gamma}^5 [\rm{MeV}]
B(E2) [\rm{e^2 fm^4}] \label{gl1}
\end{equation}
and
\begin{equation}
\Gamma_{\rm M1} [\rm{eV}] = 1.16 \cdot 10^{-2} E_{\gamma}^3 [\rm{MeV}]
B(M1) [\mu_N^2]\hspace{5mm} \quad . \label{gl2}
\end{equation}
The resonant reaction rate for an isolated narrow resonance can be
expressed in terms of the resonance strength $\omega\gamma$
(in units eV)\,\cite{wie82,her95}
\begin{equation}
\label{RRATE}
N_{\rm A} <\sigma v>_{\rm r} = 1.54 \cdot 10^5 \mu^{-3/2} T_9^{-3/2}
\sum_i (\omega \gamma)_i \exp{(-11.605 E_i / T_9)} \hspace{5mm}
{\rm cm}^3 {\rm mole}^{-1} {\rm s}^{-1} \quad ,
\end{equation}
where $E_i$ is in MeV and $T_9$ is the temperature in $10^9$K. The resonance strength
$\omega\gamma$ for a resonance is given by
\begin{equation}
\omega \gamma = \frac {2J+1}{2(2j_t+1)} \frac{\Gamma_{\rm n}
\Gamma_{\rm\gamma}}{\Gamma_{\rm tot}} \quad .
\end{equation}
The resonance strength has to be determined experimentally by low energy
neutron capture measurements or has to be derived from the calculated
partial widths.
\subsection{Non-resonant Reaction Contributions}
The non-resonant part of the neutron capture cross section has been calculated
using the DC model described in\,\cite{kim87,obe91,moh93}. The total cross
section $\sigma_{\rm nr}$ is determined by the direct capture transitions
$\sigma^{\rm DC}_i$ to all bound states with the single particle
spectroscopic factors $C^2 S_i$ in the final nucleus
\begin{equation}
\label{NR}
\sigma_{\rm nr} = \sum_{i} \: (C^{2} S)_i\sigma^{\rm DC}_i \quad .
\end{equation}
The DC cross sections $\sigma^{\rm DC}_i$ are
determined by the overlap of the scattering wave function
in the entrance channel, the bound-state wave function
in the exit channel and the multipole transition-operator.

In the stellar energy range considered here the non-resonant reaction
cross sections are predominantly determined by s-wave
and p-wave contributions. If the Q-value of the neutron capture
reaction is clearly higher than the neutron energy the cross section
for s-wave neutron capture
follows the $1/v$ law. Then the reaction rate is constant over the
entire temperature range \cite{FCZ67}. The s-wave contribution to the
reaction rate can then be directly determined from the thermal cross section
$\sigma_{\rm th}$,
\begin{equation}
N_A<\sigma v>_{\rm s} = N_A \times \sigma_{\rm th}v_{\rm th} \quad
{\rm cm}^3\,{\rm mole}^{-1}\,{\rm s}^{-1} \quad.
\end{equation}
The cross section for p-wave contributions is approximately proportional
to the relative velocity $v$, the reaction rate is therefore proportional
to the temperature \cite{rau94,FCZ67,wie90} and can be expressed by
\begin{equation}
N_A<\sigma v>_{\rm p} = \frac{1.08\cdot 10^8}{\sqrt{\mu}}\cdot
\frac{\sigma_{\rm p}(E)}{\sqrt{E}}\cdot T_9 \quad
{\rm cm}^3\,{\rm mole}^{-1}\,{\rm s}^{-1} \quad .
\end{equation}

For low Q-values the simple $1/v$- and $v$-law do not apply
anymore. A significant deviation can be observed if the neutron
energy is in the order of the Q-value. In this case the energy
dependence is given by (cf.~\cite{obe96} for a more
detailed discussion)
\begin{equation}
\sigma_{\rm DC}^{\rm E1}({\rm s} \to {\rm p}) \propto \frac{1}{\sqrt{E}}
\frac{(E+3Q)^2}{E+Q} \quad ,
\end{equation}
while a transition p$\to$s has the energy dependence
\begin{equation}
\sigma_{\rm DC}^{\rm E1}({\rm p} \to {\rm s}) \propto \frac{\sqrt{E}}{E+Q}
\quad .
\end{equation}
If $E \ll Q$ the conventional energy dependence is recovered.
From the above equations we obtain contributions to the
reaction rate which are not constant (for s-wave capture)
or proportional to $T_9$ (for p-wave capture) in the case
of small Q-values.

Several transitions considered in this work have Q-values
less than 1\,MeV (e.g., all transitions of the reaction
$^{16}$C(n,$\gamma$)$^{17}$C). For these cases the neutrons are very
loosely bound and the bound state wave functions reach out
very far. We therefore call the deviations from the
conventional energy dependence halo effects.

In the reaction $^{13}$C(n,$\gamma$)$^{14}$C we also have a 
contribution from an incoming d-wave. The energy dependence of
d-wave capture is, if $E \ll Q$, proportional to $E^{3/2}$.

We now parameterize the total non-resonant reactions
as a function of temperature $T_9$,
\begin{equation}
N_{\rm A} <\sigma v>_{\rm nr}  = A + B T_9 - C T_9^D \hspace{5mm}
{\rm cm}^3 {\rm mole}^{-1} {\rm s}^{-1} \quad .
\end{equation}
The first term and the second term arise from the
s-wave and p-wave contribution, respectively (cf.~above).
The halo effects and the d-wave contributions can be fitted 
as $C T_9^D$ with only a small error.                          
\subsection{Interferences}
If the widths of the resonances are broad an interference term has to
be added. The total cross section is then given by~\cite{rol74}
\begin{equation}
\label{eq-int}
\sigma (E) = \sigma_{\rm nr} (E) + \sigma_{\rm r} (E) +
2 \left[\sigma_{\rm nr} (E) \sigma_{\rm r} (E) \right]^{1/2}
{\rm cos}[\delta_{\rm r} (E)] \quad .
\end{equation}
In this equation $\delta_{\rm r} (E)$ is the resonance phase shift
given by
\begin{equation}
\delta_{\rm r} (E) = {\rm arctan}~\frac{\Gamma (E)}{2(E-E_{\rm r})}
\quad .
\end{equation}

Only the contributions with the same angular momentum of the
incoming wave interfere in Eq.~\ref{eq-int}. In the capture
reactions considered in this paper we find only one case
where the interference between a resonance and a direct
capture mechanism should be taken into account. This is the
$^{13}$C(n,$\gamma$)$^{14}$C reaction where the p-wave
resonance at 143\,keV interferes with the p-wave contribution
of the direct capture. In all other cases the interference can
be neglected, because either the resonance is too narrow, the
resonance energy is too high or the angular momentum of
the incoming partial waves
of the resonant and direct capture contribution differ.

With this additional term in the cross section we have to add
new terms to the reaction rate. We find that for our case of
$^{13}$C(n,$\gamma$)$^{14}$C the reaction rate
resulting from numerical integration of Eq.~\ref{eq-int} can be
described by adding an interference term
\begin{equation}
\label{eq-trate}
N_{\rm A} <\sigma v>_{\rm int}  = E T_9 + F T_9^2 \hspace{5mm}
{\rm cm}^3 {\rm mole}^{-1} {\rm s}^{-1} \quad .
\end{equation}
\subsection{Nuclear Model Input Parameter}
For the calculation of the single particle amplitude in both the
resonant as well as non-resonant neutron capture cross sections the
spectroscopic factors have to be known. These can be obtained
experimentally from single particle transfer reaction studies.
For example, the spectroscopic factors necessary for calculating
A(n,$\gamma$)B can be extracted from the reaction A(d,p)B. The
$\gamma$-widths can be extracted from reduced electromagnetic transition
strengths. For unstable nuclei where only limited or even no
experimental information is available, the spectroscopic factors and electromagnetic
transition strengths can also be calculated using nuclear
structure models like the shell model (SM).

The most important ingredients in the potential model are the wave functions
for the scattering and bound states in the entrance and exit channels.
This is the case for the DC cross sections $\sigma^{\rm DC}_i$ in
Eq.~\ref{NR} as well as for the calculation of the single-particle width
$\Gamma_{\rm s.p.}$ in Eq.~\ref{SF}.
For the calculation of these wave functions we use real folding potentials
which are given by\,\cite{obe91,kob84}
\begin{equation}
\label{FO}
V(R) = \lambda\,V_{\rm F}(R)
= \lambda\,\int\int \rho_a({\bf r}_1)\rho_A({\bf r}_2)
v_{\rm eff}\,(E,\rho_a,\rho_A,s)\,{\rm d}{\bf r}_1{\rm d}{\bf r}_2 \quad ,
\end{equation}
with $\lambda$ being a potential strength parameter close
to unity, and $s = |{\bf R} + {\bf r}_2 - {\bf r}_1|$,
where $R$ is the separation of the centers of mass of the
projectile and the target nucleus.
The density can been derived from measured
charge distributions\,\cite{vri87} or from nuclear structure models (e.g.
Hartree-Fock calculations) and the effective nucleon-nucleon
interaction $v_{\rm eff}$
has been taken in the DDM3Y parameterization\,\cite{kob84}.
The imaginary part of the potential
is very small because of the small flux into other reaction channels
and can be neglected in most cases involving neutron capture
by neutron-rich target nuclei.

For the calculation of the bound state wave function the parameter
$\lambda$ is determined from the binding energy. In the scattering
channel we try to fit $\lambda$ to reproduce the thermal elastic
scattering cross section \cite{sea92}.
In general, we assume that $\lambda$ is independent of both the
parity and the channel spin. In a few cases, however, the incoherent
scattering cross section is not negligible. Then we need to distinguish
between the different possible channel spins and determine $\lambda$
for each channel spin. One example for this is the neutron scattering
on $^{13}$C where different strength parameters are obtained due to
the known incoherent scattering data.

For the calculation of the neutron
widths of resonance states the parameter $\lambda$ is obtained
from the resonance energies. In a few cases the neutron widths
calculated with Eq.~\ref{SF} can be compared with experimental
data. This is done in Table~\ref{tab-width}. In all cases the widths
agree within a factor of 2.

Detailed shell model calculations had to be performed to
calculate the unknown excitation energies, spectroscopic factors and
electromagnetic transition rates. The code OXBASH
\cite{bro84} was used for this purpose. In most
cases we need to consider a combined
p- and sd-shell for the neutron-rich C-, N- and O-isotopes.
The wave functions of these isotopes are calculated with the
interaction WBN from Ref.~\cite{war92}.
We included no more than 1p1h-excitations from the
p- into the sd-shell. This means that for an isotope with neutron
number $N$ the occupation number of the sd-shell is $(N-8)$ for normal
parity states and $(N-7)$ for nonnormal parity states.
For higher excitations the shell model
dimensions become prohibitively high.
\section{Neutron capture by C-, N- and O-isotopes}
Most of the neutron capture rates discussed here have been calculated
previously in Ref.~\cite{rau94}. However, the predicted rates were
handicapped by several shortcomings, mainly in the determination of the
nuclear structure input parameters.
First, only limited experimental and theoretical information
was available about the single particle spectroscopic factors.
For many unstable nuclei only rather crude estimates of the spectroscopic
factors were used. Second, the electromagnetic transition strengths were not
known explicitly and systematic estimates were employed. Third,
for some of the unstable isotopes no experimental levels were known
above the threshold. Therefore, no resonant contributions were
included for these reactions.

Using the shell model we are able to calculate spectroscopic factors,
electromagnetic transition strengths, and the resonance parameters
for all reactions. While it has to be admitted that the reliability of the
shell model calculations decreases when we approach the dripline, we
nevertheless believe that our calculations represent an important improvement
compared to the previous attempts.

In the following we will discuss the neutron capture rates separately.
The parameters for the direct capture are
listed in Tables \ref{tab-dcc}, \ref{tab-dcn} and \ref{tab-dco}.
The parameterizations of the direct capture contribution to the
reaction rate is given in Table \ref{tab-rates}.
The resonance parameters are listed in
Tables \ref{tab-resc}, \ref{tab-resn} and \ref{tab-reso}.

In Figs.~\ref{fig-c13}, \ref{fig-n15} and \ref{fig-o18} we show the cross sections
determined with the help of our hybrid model for $^{13}$C(n,$\gamma$)$^{14}$C,
$^{15}$N(n,$\gamma$)$^{16}$N, and $^{18}$O(n,$\gamma$)$^{19}$O together
with the experimental data from ~\cite{ram90,shi96,mei96a,mei96b}.
We also compare our reaction rates with the previously proposed ones
in Figs.~\ref{fig-c}, \ref{fig-n} and \ref{fig-o}. In
most cases the previous theoretical rates are taken from Ref.~\cite{rau94},
except for the reactions
$^{13}$C(n,$\gamma$)$^{14}$C
$^{15}$N(n,$\gamma$)$^{16}$N, and $^{18}$O(n,$\gamma$)$^{19}$O where we
show the comparison with the experimental rates from
Refs.~\cite{ram90,shi96,mei96a,mei96b}. In Figs.~\ref{fig-smc}, \ref{fig-smn}
and \ref{fig-smo} we compare our reaction rates
with the reaction rates determined by the
Hauser-Feshbach model \cite{CTT91}.
\subsection{$^{13}$C(n,$\gamma$)$^{14}$C}
The low energy reaction cross section of $^{13}$C(n,$\gamma$)$^{14}$C
is determined by a 2$^+$ p-wave resonance at E$_{\rm n}^{\rm cm}$ = 143\,keV
which decays predominantly to the fifth excited state in $^{14}$C at
$E_{\rm x}$=7.01\,MeV (0$^-$) and by the non-resonant s-wave direct capture
to the ground state and the second excited state at E$_{\rm x}$=6.59\,MeV
(0$^+$). Additional p-wave direct capture contribution yields from the
transition to the first excited state at E$_{\rm x}$=6.09\,MeV (1$^-$).
Both the resonant cross section as well as 
the non-resonant cross sections have been measured recently~\cite{ram90,shi96}.
The experimental data indicate that the total cross section is dominated
by the p-wave resonant contribution at energies above 20\,keV. The non-resonant
contribution is considerably lower and agrees well with the value
extrapolated by the 1/v-law from the thermal cross section. 

To interpret the recent observational results in the framework of the model
described above, we determine two potential strengths from the
coherent and incoherent thermal scattering cross section. With
these strengths we calculate the s-wave capture cross section in the
channel spin formalism. The resulting theoretical thermal capture cross of
6.33\,mb is larger than the experimental absorption cross section of 1.37\,mb.
This might partially be due to the E1 polarization induced by the
incident neutron in the target nucleus~\cite{zim70}. In view of the
short-range character of the nuclear forces this polarization effect is
only important if the capture reaction takes place inside the nucleus.
This is the case for the s$\to$p transition in $^{13}$C(n,$\gamma$)$^{14}$C.
Therefore we have extrapolated the experimental thermal cross section
with an $1/v$-behavior.
Other direct transitions in this reaction are incoming p-waves where the
main contributions to direct capture come from the nuclear exterior
so that the polarization effect is supposed to be small.

At temperatures above 0.3\,GK the reaction rate is dominated by
the 143\,keV resonance, at lower temperatures p-wave
contributions determine the rate at $\approx$~0.1\,GK and s-wave
contributions at temperatures $\le$~0.05\,GK.
Since the resonance is relatively broad we cannot neglect the interference
between the resonance and the p-wave direct capture. The interference is
constructive at energies lower than the resonance energy and destructive at
higher energies.

In Fig.~\ref{fig-c13} the total cross section is shown
with the various contributions. At higher temperatures the d-wave become
important. The agreement with the experimental data from \cite{ram90,shi96}
is satisfactory although the energy dependence of the cross section is
somewhat different.
The comparison with
the rate calculated on the basis of experimental data \cite{ram90,shi96}
is shown in Fig.~\ref{fig-c}. Experimental cross section data are only available
in the energy range up to about 60\,keV. It would be interesting to measure the cross sections
at higher energies to study the interference effects with the resonance at 143\,keV and
compare them with our calculation.
The rate is considerably
lower than predicted by the Hauser-Feshbach code SMOKER as can be seen
in Fig.~\ref{fig-smc}.
This can be explained by the low level density in the compound nucleus $^{14}$C.
\subsection{$^{14}$C(n,$\gamma$)$^{15}$C}
Due to the absence of low-lying resonances the reaction rate is
given by the DC contribution only. The DC transition into the
ground state of $^{15}$C is dominant. The high spectroscopic factor
for this transition results in a clearly higher reaction rate.
The reaction rate was calculated recently by Mengoni et al.~\cite{men96}.
He compares the Maxwellian averaged cross section with 
the measurement from Ref.~\cite{bee92} and another calculation from
Ref.~\cite{wie90}.

In Table~\ref{tab-maxav} we compare our results with these previous
cross sections. Compared to both other calculations our result of
10.14\,$\mu$b is slightly larger. All calculations are clearly
larger than the experimental data. The reason for this
discrepancy is still unknown. There are plans to remeasure the
cross section.
\subsection{$^{15}$C(n,$\gamma$)$^{16}$C}
This reaction is also purely direct since there exists no resonance level
near the threshold \cite{fir96}. The direct capture proceeds to the $0^+$ ground
state, the first excited $2^+$ state at 1.766\,MeV and the second
excited $0^+$ at 3.027\,MeV. Since our spectroscopic factors -- especially
for the transition to the second excited state -- are
higher than estimated in Ref.~\cite{rau94}, the new reaction rate is
larger. The new rate is also considerably lower than the predicted
value by SMOKER (Fig.~\ref{fig-smc}).
\subsection{$^{16}$C(n,$\gamma$)$^{17}$C}
The reaction rate is dominated by the direct transition to the
second excited $1/2^+$ state at 0.295\,MeV. Again the spectroscopic
factor and the reaction rate are higher than estimated by Rauscher et al.
\cite{rau94} (Fig.~\ref{fig-c}) but seems to be in good agreement with
the Hauser-Feshbach predictions (see Fig.~\ref{fig-smc}).
The contribution of the resonance at 440\,keV is negligible.
This resonance was recently identified
experimentally by Raimann et al. \cite{rai96}. The Maxwellian
averaged cross section at 30\,keV was calculated by Mengoni. In
Table~\ref{tab-maxav} we compare our results. The cross sections differ
only by about 10 per cent.
\subsection{$^{17}$C(n,$\gamma$)$^{18}$C}
From the shell model calculations we obtain four resonances
in the energy range between $E_{\rm r} = 684$\,keV and
$E_{\rm r} = 796$\,keV. The reaction rate is only slighty dependent on these resonances,
because the resonance energies are more than 700 keV above threshold
For lower temperatures
the direct capture dominates. This rate has
not been calculated previously, yet the calculated rate is in reasonable
agreement with the prediction by the Hauser-Feshbach model.
This agreement seems to be a fortuitous circumstance.
\subsection{$^{15}$N(n,$\gamma$)$^{16}$N}
The reaction has been investigated at energies between
25\,keV and 400\,keV recently~\cite{mei96a}. The cross section is clearly dominated
by p-wave transitions for energies below 500\,keV.
The s-wave contribution, which is obtained from an extrapolation of
the thermal absorption cross section, is negligible at
thermonuclear energies.

In Fig.~\ref{fig-n15} the cross section is compared with the
measured data points from Ref.~\cite{mei96a}. The agreement is
very good.
The resulting reaction rate is in good agreement
with previous calculations~\cite{rau94}. The present calculation is
also very similar to the calculations in Ref.~\cite{mei96a}.
The small difference is due to the use of a folding potential in this
work.
\subsection{$^{16}$N(n,$\gamma$)$^{17}$N}
Several unbound levels in $^{17}$N are known from transfer and
$\beta$-delayed neutron decay studies~\cite{TWC93,SGV95}. However,
no spin assignment is available for the states in the stellar
energy range. While the shell model predicts several levels in this
energy region the available information does not allow identification
of the levels with their experimental counterparts. There\-fore we
used the shell model energies in our calculation for the resonant
contribution. In Ref.~\cite{rau94} the experimental energy of the
198\,keV resonance was used with a hypothetical $5/2^+$ assignment.

The difference in resonance energies in the previous and the present
calculations explains the ratio of the rates. Moreover, the
inclusion of the direct transition to a $5/2^-$ state at 4.415\,MeV
with a spectroscopic factor close to unity leads to an enhancement of
the reaction rate (see Fig.~\ref{fig-n}). Yet the rate is
considerably lower than the rate predicted by the Hauser-Feshbach
model as shown in Fig.~\ref{fig-smn}.
\subsection{$^{17}$N(n,$\gamma$)$^{18}$N}
Several bound states in $^{18}$N have been identified in transfer,
charge exchange, and $^{18}$C $\beta$-decay
studies \cite{TWC95}. The measurement of the $\beta$-delayed neutron
decay of $^{18}$C \cite{SGV95} yields information about some neutron
unbound states, but the identified levels are too high in excitation
energy to be of relevance for the neutron capture process discussed here.
Due to the lack of experimental information about potential resonance
levels its contribution was calculated on the basis of shell model
predictions for level energies and single particle strengths.
The strongest contribution to its direct capture components
is the transition to a $1^-$ state at 1.165\,MeV with a spectroscopic
factor of 0.7. This transition was not included in the previous
calculation \cite{rau94}.

Our rate is more than one order of magnitude larger than the previous
estimate~\cite{rau94} but seems to agree reasonably well with the
Hauser-Feshbach calculation.
\subsection{$^{18}$N(n,$\gamma$)$^{19}$N}
Only very limited experimental information is available about the
level structure of $^{19}$N \cite{TWC95}. Multi-particle transfer studies
identified some of the excited bound states and helped to determine the
corresponding excitation energies. In the study of the $\beta$-delayed
neutron emission of $^{19}$C \cite{ORB96} several
neutron unbound levels above 6.3\,MeV were observed. No states, however,
were identified near the neutron threshold of 5.32~MeV. We therefore
rely in our estimate on shell model predictions only.   
A resonance a few keV above the threshold dominates the reaction
rate for temperatures below $T_9=1$. According to the factor
$\exp{(-11.605 E_i / T_9)}$ in Eq.~\ref{RRATE} the resonant
reaction rate varies by about a factor of 2 when assuming an
uncertainty of the resonance energy from 0-50\,keV.
According to the factor $\exp{(-11.605 E_i / T_9)}$ in Eq.~\ref{RRATE}
the resonant reaction rate varies by about a factor of 2 at $T_9=1$ when
assuming an uncertainty of the resonance energy from 0-50\,keV.

In this temperature range our rate
is therefore two to four orders of magnitude larger
compared to Ref.~\cite{rau94}. For higher temperatures other resonances
become important which is indicated by the agreement with the Hauser-Feshbach 
estimate for higher temperatures (Fig.~\ref{fig-smn}).
\subsection{$^{18}$O(n,$\gamma$)$^{19}$O}
The cross section for this reaction was recently measured over a 
wide energy range between 25\,keV and 400\,keV \cite{mei96b}.
The theoretical analysis took into account possible s- and p-wave
transitions to the ground state and the first excited states as well
as contributions from previously observed but unpublished
higher energy resonances \cite{MDN81}.
While the present evaluation is based on the same resonance parameters,
the direct capture cross section has been reevaluated to improve the
extrapolation of the data over the entire energy range. The present
reaction rate differs only slightly from the previous result.
The s-wave is determined by a transition to a state a few keV
below the threshold. Neither the spin assignment nor the energy are
known exactly. We extrapolate the experimental thermal cross section
with an $1/v$-behavior to determine the s-wave contribution.
In Fig.~\ref{fig-o18} we compare the calculated cross section
with the experimental data from Ref.~\cite{mei96b}. The data points
at weighted neutron energies of 150\,keV and 370\,keV are clearly
enhanced.

The calculated reaction rate agrees very well with the calculation
of Ref.~\cite{mei96b}. Again the small difference is primarily due
to the use of a folding potential.
\subsection{$^{19}$O(n,$\gamma$)$^{20}$O}
Several excited states -- two of them forming a doublet at 7.622\,MeV
excitation energy -- are known in the compound nucleus $^{20}$O
above the neutron threshold at 7.608~MeV\,\cite{ajz87,TCG97}.
These states could be identified with shell model levels.
From the shell model we also obtain a number of additional resonance
states. The previous reaction rate included only two resonances. For high
temperatures our rate is therefore about one order of magnitude larger
than the previous estimate (Fig.~\ref{fig-o}) but seems to be in
good agreement with the Hauser-Feshbach calculation (Fig.~\ref{fig-smo}).
\subsection{$^{20}$O(n,$\gamma$)$^{21}$O}
The shell model predicts for the first excited state above the
$5/2^+$ ground state of $^{21}$O a spin and parity of $1/2^+$ 
and an excitation energy of 1.33\,MeV. With a high spectroscopic factor
of 0.811 the direct p $\to$ s transition to this state
dominates the reaction rate. Previously only the transition to
the ground state was taken into account. Therefore the new rate is
higher by a factor of approximately 5. The contributions of the
two resonances are small.
\subsection{$^{21}$O(n,$\gamma$)$^{22}$O}
The new reaction rate is between two and three orders of magnitude
higher. The spectroscopic factor for the transition to the
ground state is 5.222, much higher than 0.2, which was estimated
in Ref.~\cite{rau94}. Still, the transitions to two excited
states are larger than the ground state transition since they are
p $\to$ s transitions. Moreover, the resonance at 14\,keV contributes
at low temperatures.
\section{Summary and Discussion}
CNO-isotopes can act as neutron poisons in s-process environments like
red giants or asymptotic giant branch stars. They might also serve as pathways 
to the production of heavier nuclei in inhomogeneous big bang environments.
In either case, high accuracy cross sections and reaction rates are strongly needed
for relevant and precise nucleosynthesis predictions. 

In the present paper we combine the latest experimental information based on cross section
measurements (for stable target nuclei) and on indirect measurements of transfer, charge exchange
and decay processes (for radioactive target nuclei) with theoretical direct capture and shell model
calculations to derive reliable rates for neutron capture on neutron-rich C-, N-, and O-isotopes. 
Thus, the reaction rates provided here contain the latest information available and should replace the 
corresponding rates given in previous compilations~\cite{rau94,TSO93,OKB97}. 

\begin{table}[htb]
\caption{Comparison of the calculated neutron widths of
resonance states with experimental values.}
\begin{center}
\renewcommand{\arraystretch}{0.7}
\begin{tabular}{|rrrrrr|}
reaction & $E_{\rm x}$ & $J^{\pi}$ &
$E_{\rm res}$[MeV] & $\Gamma_{\rm n}$ (calc) [keV] &
$\Gamma_{\rm n}$ (exp) [keV] \\
\hline
$^{13}$C(n,$\gamma$)$^{14}$C & 8.320 & $2^+$ & 0.143 & 3.15 & $3.4 \pm 0.6$
\tablenotemark[1] \\
\hline
$^{15}$N(n,$\gamma$)$^{16}$N & 3.360 & $1^+$ & 0.869 & 21.85 & $15 \pm 5$ 
\tablenotemark[2] \\
\hline
$^{18}$O(n,$\gamma$)$^{19}$O & 4.109 & $3/2^+$ & 0.152 & 0.05 & $< 15$
\tablenotemark[3] \\
& 4.328 & $5/2^-$ & 0.371 & 6 $\times 10^{-4}$ & $< 15$ \tablenotemark[3] \\
& 4.582 & $3/2^-$ & 0.625 & 28.5 & $52 \pm 3$ \tablenotemark[3] \\
& 4.703 & $5/2^+$ & 0.746 & 0.04 & $< 15$ \tablenotemark[3] \\
\end{tabular}
\tablenotetext[1]{from Ref.~\cite{ajz86a}}
\tablenotetext[2]{from Ref.~\cite{ajz86b}}
\tablenotetext[3]{from Ref.~\cite{mei96b}}
\end{center}
\label{tab-width}
\end{table}

\begin{table}[htb]
\caption{Considered transitions for the direct capture reactions on
C-isotopes.
Transitions with very small contributions are not included in the
table. The spectroscopic factors are from shell model calculations
unless stated otherwise. The s-wave transitions for the 
reaction $^{13}$C(n,$\gamma)^{14}$C were obtained by extrapolating the 
thermal absorption cross section. Therefore the spectroscopic factors 
for these transitions are not listed.} 
\begin{center} 
\renewcommand{\arraystretch}{0.7}
\begin{tabular}{|rrrrrr|}
reaction & Q-value (MeV) & $J^{\pi}$ & $E_x$ (MeV) & transition & $C^2 S$ \\
\hline
$^{13}$C(n,$\gamma)^{14}$C & 8.176 & $0^+$ & 0.000 & s$\to$1p$_{1/2}$ & \\
 & & & & d$\to$1p$_{1/2}$ & 1.734 \\
 & & $1^-$ & 6.094 & p$\to$2s$_{1/2}$ & 0.750 \tablenotemark[1] \\
 & & $0^+$ & 6.589 & s$\to$2p$_{1/2}$ & \\
 & & $3^-$ & 6.728 & p$\to$1d$_{5/2}$ & 0.650 \tablenotemark[1] \\
 & & $2^+$ & 7.012 & s$\to$2p$_{3/2}$ & \\
 & & $2^-$ & 7.341 & p$\to$1d$_{5/2}$ & 0.720 \tablenotemark[1] \\
$^{14}$C(n,$\gamma)^{15}$C & 1.218 & $1/2^+$ & 0.000 & p$\to$2s$_{1/2}$ &
0.980 \\ 
 & & $5/2^+$ & 0.740 & p$\to$1d$_{5/2}$ & 0.943 \\
$^{15}$C(n,$\gamma)^{16}$C & 4.251 & $0^+$ & 0.000 & p$\to$2s$_{1/2}$ &
0.601 \\
 & & $2^+$ & 1.766 & p$\to$1d$_{5/2}$ & 0.493 \\
 & & $0^+$ & 3.027 & p$\to$2s$_{1/2}$ & 1.344 \\
$^{16}$C(n,$\gamma)^{17}$C & 0.729 & $3/2^+$ & 0.000 & p$\to$1d$_{3/2}$ &
0.035 \\
 & & $5/2^+$ & 0.032 & p$\to$1d$_{5/2}$ & 0.701 \\
 & & $1/2^+$ & 0.295 & p$\to$2s$_{1/2}$ & 0.644 \\
$^{17}$C(n,$\gamma)^{18}$C & 4.180 & $0^+$ & 0.000 & p$\to$1d$_{3/2}$ &
0.103 \\
 & & $2^+$ & 2.114 & p$\to$1d$_{5/2}$ & 1.081 \\
 & & & & p$\to$2s$_{1/2}$ & 0.015 \\
 & & $2^+$ & 3.639 & p$\to$2s$_{1/2}$ & 0.525 \\
\end{tabular}
\tablenotetext[1]{from Ref.~\cite{ajz86a}}
\end{center}
\label{tab-dcc}
\end{table}

\begin{table}[htb]
\caption{Considered transitions for the direct capture reactions on
N-isotopes.
Transitions with very small contributions are not included in the
table. The spectroscopic factors are from shell model calculations
unless stated otherwise. The s-wave transitions for the reaction
$^{15}$N(n,$\gamma)^{16}$N
were obtained by extrapolating the thermal
absorption cross section. These transitions are not listed in the 
table.}
\begin{center}
\renewcommand{\arraystretch}{0.7}
\begin{tabular}{|rrrrrr|}
reaction & Q-value (MeV) & $J^{\pi}$ & $E_x$ (MeV) & transition & $C^2 S$ \\
\hline
$^{15}$N(n,$\gamma)^{16}$N & 2.491 & $2^-$ & 0.000 & p$\to$1d$_{5/2}$ &
0.550 \tablenotemark[1] \\
 & & $0^-$ & 0.120 & p$\to$2s$_{1/2}$ & 0.460 \tablenotemark[1] \\
 & & $3^-$ & 0.298 & p$\to$1d$_{5/2}$ & 0.540 \tablenotemark[1] \\
 & & $1^-$ & 0.397 & p$\to$2s$_{1/2}$ & 0.520 \tablenotemark[1] \\
$^{16}$N(n,$\gamma)^{17}$N & 5.883 & $5/2^-$ & 1.907 & p$\to$1d$_{5/2}$ &
0.207 \\
 & & $7/2^-$ & 3.129 & p$\to$1d$_{5/2}$ & 1.457 \\
 & & $5/2^-$ & 4.415 & p$\to$2s$_{1/2}$ & 0.921 \\
$^{17}$N(n,$\gamma)^{18}$N & 2.825 & $2^-$ & 0.121 & p$\to$1d$_{5/2}$ &
0.700 \\
 & & $3^-$ & 0.747 & p$\to$1d$_{5/2}$ & 0.689 \\
 & & $1^-$ & 1.165 & p$\to$2s$_{1/2}$ & 0.705 \\
$^{18}$N(n,$\gamma)^{19}$N & 5.328 & $3/2^-$ & 1.682 & p$\to$1d$_{5/2}$ &
0.579 \\
 & & $1/2^+$ & 2.115 & s$\to$2p$_{3/2}$ & 0.001 \\
 & & & & s$\to$2p$_{1/2}$ & 0.0007 \\
 & & $5/2^-$ & 2.173 & p$\to$1d$_{5/2}$ & 0.354 \\
 & & $5/2^+$ & 2.375 & s$\to$2p$_{3/2}$ & 0.001 \\
 & & $3/2^-$ & 3.591 & p$\to$2s$_{1/2}$ & 0.461 \\
 & & $3/2^+$ & 3.799 & s$\to$2p$_{3/2}$ & 0.007 \\
 & & & & s$\to$2p$_{1/2}$ & 0.001 \\
 & & $1/2^-$ & 4.126 & p$\to$2s$_{1/2}$ & 0.606 \\
 & & $3/2^-$ & 4.438 & p$\to$2s$_{1/2}$ & 0.035 \\
 & & $1/2^-$ & 5.101 & p$\to$2s$_{1/2}$ & 0.440 \\
 & & $5/2^+$ & 5.130 & s$\to$2p$_{3/2}$ & 0.001 \\
 & & $3/2^-$ & 5.173 & p$\to$2s$_{1/2}$ & 0.097 \\
\end{tabular}
\tablenotetext[1]{from Ref.~\cite{boh72}}
\end{center}
\label{tab-dcn}
\end{table}

\begin{table}[htb]
\caption{Considered transitions for the direct capture reactions
on O-isotopes. 
Transitions with very small contributions are not included in the
table. The spectroscopic factors are from shell model calculations
unless stated otherwise. The s-wave transitions for the reaction
$^{18}$O(n,$\gamma)^{19}$O were obtained by extrapolating the thermal
absorption cross section. Therefore the spectroscopic factors for
these transitions are not listed.}
\begin{center}
\renewcommand{\arraystretch}{0.7}
\begin{tabular}{|rrrrrr|}
reaction & Q-value (MeV) & $J^{\pi}$ & $E_x$ (MeV) & transition & $C^2 S$ \\
\hline
$^{18}$O(n,$\gamma)^{19}$O & 3.957 & $5/2^+$ & 0.000 & p$\to$1d$_{5/2}$ &
0.570 \tablenotemark[1] \\ 
 & & $1/2^+$ & 1.472 & p$\to$2s$_{1/2}$ & 1.000 \tablenotemark[1] \\
 & & $3/2^-$ & 3.945 & s$\to$2p$_{3/2}$ & \\
$^{19}$O(n,$\gamma)^{20}$O & 7.606 & $0^+$ & 0.000 & p$\to$1d$_{5/2}$ &
3.427 \\
 & & $2^+$ & 1.674 & p$\to$1d$_{5/2}$ & 0.731 \\
 & & & & p$\to$2s$_{1/2}$ & 0.142 \\
 & & $4^+$ & 3.570 & p$\to$1d$_{5/2}$ & 1.021 \\
 & & $2^+$ & 4.072 & p$\to$2s$_{1/2}$ & 0.573 \\
 & & $3^+$ & 5.447 & p$\to$2s$_{1/2}$ & 0.817 \\
$^{20}$O(n,$\gamma)^{21}$O & 3.806 & $5/2^+$ & 0.000 & p$\to$1d$_{5/2}$ & 
0.345 \\
 & & $1/2^+$ & 1.330 & p$\to$2s$_{1/2}$ & 0.811 \\
$^{21}$O(n,$\gamma)^{22}$O & 6.850 & $0^+$ & 0.000 & p$\to$1d$_{5/2}$ &
5.222 \\
 & & $2^+$ & 3.374 & p$\to$2s$_{1/2}$ & 0.822 \\
 & & $3^+$ & 4.830 & p$\to$2s$_{1/2}$ & 0.771 \\
\end{tabular}
\tablenotetext[1]{from Ref.~\cite{ajz87}}
\end{center}
\label{tab-dco}
\end{table}

\begin{table}[htb]
\caption[direct capture parameters ]
{Parameters for the direct capture reactions. The interference term
with the parameters $E$ and $F$ is only used for the reaction
$^{13}$C(n,$\gamma$)$^{14}$C.}
\begin{center}
\begin{tabular}{|l|c|r|r|r|r|r|}
 & \multicolumn{1}{c}{$A$}& \multicolumn{1}{c}{$B$} & 
  \multicolumn{1}{c}{$C$} & \multicolumn{1}{c}{$D$} &
  \multicolumn{1}{c}{$E$} & \multicolumn{1}{c|}{$F$}
\\ \hline\hline
${\rm ^{13}C(n,\gamma)^{14}C}$ & $182.310$ & $3296.558$ & $-5534.721$ &
 $2.009$ & $1794.927$ & $-1103.269$ \\ 
${\rm ^{14}C(n,\gamma)^{15}C}$ & --- & $4754.286$ & $752.370$ &
 $1.630$ &&\\ 
${\rm ^{15}C(n,\gamma)^{16}C}$ & --- & $2637.750$ & $304.878$ &
 $1.644$ &&\\
${\rm ^{16}C(n,\gamma)^{17}C}$ & --- & $2861.539$ & $1166.516$ &
 $1.311$ &&\\
${\rm ^{17}C(n,\gamma)^{18}C}$ & --- & $1334.578$ & $337.957$ &
 $1.472$ &&\\
${\rm ^{15}N(n,\gamma)^{16}N}$ & 3.18 & $3783.415$ & $335.198$ &
 $1.716$ &&\\ 
${\rm ^{16}N(n,\gamma)^{17}N}$ & --- & $3649.913$ & $437.549$ &
 $1.633$ &&\\ 
${\rm ^{17}N(n,\gamma)^{18}N}$ & --- & $3417.690$ & $358.029$ &
 $1.660$ &&\\ 
${\rm ^{18}N(n,\gamma)^{19}N}$ & $13.838$ & $4051.118$ & $966.727$ &
 $1.412$ &&\\ 
${\rm ^{18}O(n,\gamma)^{19}O}$ & $21.357$ & $8300.018$ & $596.897$ &
 $1.770$ &&\\
${\rm ^{19}O(n,\gamma)^{20}O}$ & --- & $7275.368$ & $432.266$ &
 $1.747$ &&\\
${\rm ^{20}O(n,\gamma)^{21}O}$ & --- & $6474.727$ & $493.516$ &
 $1.750$ &&\\ 
${\rm ^{21}O(n,\gamma)^{22}O}$ & --- & $7327.938$ & $543.151$ &
 $1.747$ &&\\
\end{tabular}
\end{center}
\label{tab-rates}
\end{table}             

\begin{table}[bh]
\caption[Resonance parameters 1]
{Adopted values for the resonance parameters for capture reactions 
on carbon isotopes. The 
neutron and $\gamma$ widths
are calculated as explained in the text unless stated otherwise.}
\begin{center}
\renewcommand{\arraystretch}{0.7}
\begin{tabular}{|l|c|c|c|c|c|c|}
Reaction & $E_{\rm x}$ & $J^{\pi}$ & 
$E_{\rm res}$ & $\Gamma_{\rm n}$ & 
$\Gamma_{\gamma}$ & $\omega\gamma$\\
& $[$\footnotesize MeV$]$  & & $[$\footnotesize MeV$]$ & 
$[$\footnotesize eV$]$ & $[$\footnotesize eV$]$ & $[$\footnotesize eV$]$ \\ 
\hline
\hline
$^{13}\mbox{C(n,}\gamma\mbox{)}$$^{14}\mbox{C}\quad$ &
8.320 & $2^{+}$ & 0.143 & 3$\,$400 \tablenotemark[1] & 0.215 
\tablenotemark[2] & 0.269  \\ 
$^{16}\mbox{C(n,}\gamma\mbox{)}$$^{17}\mbox{C}\quad$ &
1.180 & $1/2^{-}$ & 0.451 & $\,\enspace$950 & 2.82$\,\cdot\, 10^{-4}$ & 
2.82$\,\cdot\, 10^{-4}$ \\ 
$^{17}\mbox{C(n,}\gamma\mbox{)}$$^{18}\mbox{C}\quad$ &
4.864 & $4^{+}$ & 0.684 & $\,\quad$50 &  6.11$\,\cdot\, 10^{-4}$ 
&6.87$\,\cdot\,10^{-4}$\\
& 4.915 & $3^{+}$ & 0.735 & 8$\,$450 & 7.37$\,\cdot\, 10^{-3}$ 
&6.45$\,\cdot\,10^{-3}$\\
& 4.972 & $1^{-}$ & 0.792 & 5$\,$440 & 5.85$\,\cdot\, 10^{-2}$ &0.022\\
& 4.976 & $2^{+}$ & 0.796 & 8$\,$510 & 1.85$\,\cdot\, 10^{-3}$ 
&1.16$\,\cdot\,10^{-3}$\\
\end{tabular}
\tablenotetext[1]{from Ref.~\cite{ajz86a}}
\tablenotetext[2]{from Ref.~\cite{ram90}}
\end{center}
\label{tab-resc}
\end{table}
\clearpage
\newpage

\begin{table}[htbp]
\caption[resonance parameters 2]
{Adopted values for the resonance parameters for capture reactions 
on nitrogen isotopes. The neutron and $\gamma$ widths are
calculated as explained in the text unless stated otherwise.}
\begin{center}
\renewcommand{\arraystretch}{0.7}
\begin{tabular}{|l|c|c|c|r|c|c|}
Reaction & $E_{\rm x}$ & $J^{\pi}$ & 
$E_{\rm res}$ & $\Gamma_{\rm n}\enspace\,$ & 
$\Gamma_{\gamma}$ & $\omega\gamma$\\
& $[$\footnotesize MeV$]$  & & $[$\footnotesize MeV$]$ & 
$[$\footnotesize eV$]\enspace\,$ & 
$[$\footnotesize eV$]$ & $[$\footnotesize eV$]$ \\ 
\hline
\hline
$^{15}\mbox{N(n,}\gamma\mbox{)}$$^{16}\mbox{N}\quad$ &
3.360 & $1^{+}$ & 0.862 & 15$\,$000 \tablenotemark[1] & 0.455 & 0.341  \\ 
$^{16}\mbox{N(n,}\gamma\mbox{)}$$^{17}\mbox{N}\quad$ &
5.904 & $7/2^{-}$ & 0.021 & 0.032 & 4.80$\,\cdot\,10^{-2}$ & 
0.015 \\ 
& 6.121 & $5/2^{+}$ & 0.238 & 1.2 & 4.80$\,\cdot\,10^{-2}$& 0.027\\
& 6.325 & $3/2^{+}$ & 0.442 & 20 & 5.46$\,\cdot\,10^{-2}$& 0.022\\
& 6.372 & $7/2^{+}$ & 0.489 & 20 & 1.52$\,\cdot\,10^{-2}$& 0.012\\
& 6.373 & $5/2^{+}$ & 0.490 & 600 & 0.110 & 0.066\\
& 6.470 & $1/2^{+}$ & 0.587 & 1$\,$750 & 2.510 & 0.501\\
& 6.685 & $3/2^{-}$ & 0.802 & 12$\,$500 & 5.660 & 2.263\\
& 6.737 & $7/2^{+}$ & 0.854 & 70 & 4.17$\,\cdot\,10^{-2}$&  0.033\\
& 6.835 & $3/2^{+}$ & 0.952 & 360 & 0.478& 0.191\\
$^{17}\mbox{N(n,}\gamma\mbox{)}$$^{18}\mbox{N}\quad$ &
2.875 & $3^{-}$ & 0.050 & 1.6 &  9.61$\,\cdot\, 10^{-3}$ & 0.017\\
& 2.949 & $2^{+}$ & 0.124 & 1$\,$390 & 9.72$\,\cdot\, 10^{-2}$ & 0.121\\
& 3.068 & $1^{+}$ & 0.243 & 1$\,$060 & 0.209 & 0.157\\
& 3.374 & $3^{+}$ & 0.549 & 50 & 0.107 & 0.187\\
& 3.437 & $2^{-}$ & 0.612 & 10$\,$180 & 0.517& 0.646\\
& 3.631 & $0^{+}$ & 0.806 & 2$\,$350 & 6.26$\,\cdot\,10^{-2}$& 0.016\\
& 3.644 & $1^{-}$ & 0.819 & 2$\,$370 & 0.319& 0.239\\
& 3.722 & $2^{+}$ & 0.897 & 190 & 4.48$\,\cdot\,10^{-2}$& 0.056\\
$^{18}\mbox{N(n,}\gamma\mbox{)}$$^{19}\mbox{N}\quad$ &
5.335 & $3/2^{+}$ & 0.007 & 7.8 & 0.128 & 0.050\\
& 5.479 & $3/2^{+}$ & 0.151 & 70 & 0.072 & 0.029\\
& 5.498 & $7/2^{+}$ & 0.170 & 40 & 2.40$\,\cdot\, 10^{-2}$ &0.019\\
& 5.634 & $9/2^{-}$ & 0.306 & 120 & 5.72$\,\cdot\, 10^{-4}$ 
&5.7$\,\cdot\,10^{-4}$\\
& 5.770 & $5/2^{+}$ & 0.442 & 170 & 4.28$\,\cdot\,10^{-2}$& 0.026\\
& 5.778 & $3/2^{+}$ & 0.450 & 120 & 0.320& 0.128\\
& 5.858 & $3/2^{-}$ & 0.530 & 8$\,$470 & 2.520 & 1.008\\
& 5.955 & $7/2^{+}$ & 0.627 & 180 & 8.26$\,\cdot\,10^{-2}$& 0.066\\
& 6.006 & $5/2^{+}$ & 0.678 & 190 & 4.85$\,\cdot\,10^{-2}$& 0.029\\
& 6.094 & $7/2^{-}$ & 0.766 & 9$\,$990 & 0.306& 0.245\\
& 6.125 & $7/2^{+}$ & 0.797 & 950 & 5.08$\,\cdot\,10^{-2}$& 0.041\\
& 6.128 & $3/2^{+}$ & 0.800 & 140 & 1.130 & 0.448\\
& 6.130 & $5/2^{+}$ & 0.802 & 310 & 7.86$\,\cdot\,10^{-2}$& 0.047\\
& 6.152 & $1/2^{+}$ & 0.824 & 2$\,$090 & 1.100 & 0.220\\
\end{tabular}
\tablenotetext[1]{from Ref.~\cite{ajz86b}}
\end{center}
\label{tab-resn}
\end{table}
\clearpage
\newpage

\begin{table}[b]
\caption[resonance parameters 3]
{Adopted values for the resonance parameters for capture reactions 
on oxygen isotopes. The neutron and $\gamma$ widths are calculated
as explained in the text unless stated otherwise.}
\begin{center}
\renewcommand{\arraystretch}{0.7}
\begin{tabular}{|l|c|c|c|r|c|c|}
Reaction & $E_{\rm x}$ & $J^{\pi}$ & 
$E_{\rm res}$ & $\Gamma_{\rm n}\enspace\,$ & 
$\Gamma_{\gamma}$ & $\omega\gamma$\\
& $[$\footnotesize MeV$]$  & & $[$\footnotesize MeV$]$ & 
$[$\footnotesize eV$]\enspace\,$ & 
$[$\footnotesize eV$]$ & $[$\footnotesize eV$]$ \\ 
\hline
\hline
$^{18}\mbox{O(n,}\gamma\mbox{)}$$^{19}\mbox{O}\quad$ &
4.109 & $3/2^{+}$ & 0.152 &  50 & 1.4$\,\cdot\,10^{-2}$ & 0.028 \\ 
& 4.328 & $5/2^{-}$ & 0.371 & 0.6& 0.600 & 0.900\\
& 4.582 & $3/2^{-}$ & 0.625 & 52$\,$000 \tablenotemark[1] & 1.900 & 3.800 \\
& 4.703 & $5/2^{+}$ & 0.746 & 40 & 0.500 & 1.483 \\
$^{19}\mbox{O(n,}\gamma\mbox{)}$$^{20}\mbox{O}\quad$ &
7.622 & $3^{-}$ & 0.015 & 11 & 6.52$\,\cdot\,10^{-2}$ & 0.038\\
& 7.622 &$4^{+}$ & 0.015 & 1.1&1.11$\,\cdot\,10^{-2}$&8.24$\,\cdot\,10^{-3}$ \\
& 7.638  & $4^{-}$ & 0.031 & 2.3 & 1.82$\,\cdot\,10^{-2}$ & 0.014 \\ 
& 7.646 & $2^{-}$ & 0.039 & 3.3 & 0.109 & 0.044\\
& 7.739 & $1^{-}$ & 0.132 & 550 & 12.33 & 3.015\\
& 7.754 & $4^{+}$ & 0.147 & 1$\,$190 & 0.301 & 0.226\\
& 7.855 & $5^{-}$ & 0.248 & 1.8 & 1.45$\,\cdot\,10^{-2}$ & 0.013\\
& 7.970 & $2^{+}$ & 0.363 & 2$\,$480 & 4.46$\,\cdot\,10^{-2}$ & 0.019\\
& 8.160 & $3^{-}$ & 0.553 & 2$\,$890 & 0.352 & 0.205\\
& 8.403 & $1^{-}$ & 0.796 & 9$\,$150 & 0.362 & 0.090\\
& 8.439 & $2^{-}$ & 0.832 & 2$\,$090 & 0.333 & 0.139\\
& 8.533 & $3^{-}$ & 0.926 & 8$\,$840 & 0.688 & 0.401\\
& 8.552 & $2^{-}$ & 0.945 & 1$\,$480 & 0.238 & 0.099\\
& 8.558 & $3^{+}$ & 0.951 & 20$\,$000 & 0.320 & 0.187\\
$^{20}\mbox{O(n,}\gamma\mbox{)}$$^{21}\mbox{O}\quad$ &
4.343 & $5/2^{-}$ & 0.536 & 0.8 & 0.121 & 0.316\\
&4.765 & $3/2^{-}$ & 0.958 & 5$\,$450 & 0.631 & 1.262\\
$^{21}\mbox{O(n,}\gamma\mbox{)}$$^{22}\mbox{O}\quad$ &
6.863 & $4^{+}$ & 0.014 & 0.7&6.83$\,\cdot\,10^{-4}$ & 5.12$\,\cdot\,10^{-4}$\\
& 7.357 & $4^{+}$ & 0.508 & 26$\,$810 & 2.20$\,\cdot\,10^{-3}$ & 
1.65$\,\cdot\,10^{-3}$\\
& 7.397 & $2^{-}$ & 0.548 & 780 & 0.014 &5.83$\,\cdot\,10^{-3}$\\
& 7.472 & $3^{-}$ & 0.623 & 790 & 0.023 & 0.013\\
& 7.515 & $1^{-}$ & 0.666 & 42$\,$710 & 0.146& 0.036\\
\end{tabular}
\tablenotetext[1]{from Ref.~\cite{mei96b}}
\end{center}
\label{tab-reso}
\end{table}

\begin{table}[htb]
\caption
{Comparison of the Maxwellian averaged cross section $<\sigma v> / kT$
at $kT=30$\,keV with other works.}
\begin{center}
\renewcommand{\arraystretch}{0.7}
\begin{tabular}{|rrrrr|}
reaction & this work & Ref.~\cite{men96} & Ref.~\cite{wie90} & 
Ref.~\cite{bee92} \\
$^{14}$C(n,$\gamma$)$^{15}$C & 10.14 & 8.3 & 8.4 & 1.87 $\pm$ 0.43 \\
$^{16}$C(n,$\gamma$)$^{17}$C & 4.71 & 4.3 &&\\
\end{tabular}
\end{center}
\label{tab-maxav}
\end{table}

\acknowledgments
This work was supported by
Fonds zur F\"orderung der wissenschaftlichen Forschung 
(FWF project P13246-TPH).

\newpage
\begin{figure}
\caption[fig1]{Calculated cross section of the reaction $^{13}$C(n,$\gamma$)$^{14}$C
compared with experimental data from Ref.~\cite{ram90,shi96}.}
\label{fig-c13}
\end{figure}
\begin{figure}
\caption[fig2]{Calculated cross section of the reaction $^{15}$N(n,$\gamma$)$^{16}$N
compared with experimental data from Ref.~\cite{mei96a}.}
\label{fig-n15}
\end{figure}
\begin{figure}
\caption[fig3]{Calculated cross section of the reaction $^{18}$O(n,$\gamma$)$^{19}$O
compared with experimental data from Ref.~\cite{mei96b}.}
\label{fig-o18}
\end{figure}
\begin{figure}
\caption[fig4]{Comparison of our new reaction rates for neutron
capture on C-isotopes with previous
direct capture calculations. Shown is the ratio of the new rates to
the rates published in Ref.~\cite{rau94} for all rates except
$^{13}$C(n,$\gamma$)$^{14}$C where the ratio to the rate determined
from the experimental data of
Ref.~\cite{ram90,shi96} is shown.}
\label{fig-c}
\end{figure}
\begin{figure}
\caption[fig5]{Comparison of our new reaction rates for neutron capture
on N-isotopes with previous
direct capture calculations. Shown is the ratio of the new rates to
the rates published in Ref.~\cite{rau94} for all rates except
$^{15}$N(n,$\gamma$)$^{16}$N where the ratio to the rate
determined
from the experimental data of
Ref.~\cite{mei96a} is shown.}
\label{fig-n}
\end{figure}
\begin{figure}
\caption[fig6]{Comparison of our new reaction rates for neutron capture
on O-isotopes with previous direct capture calculations. Shown is the ratio
of the new rates to the rates published in
Ref.~\cite{rau94} for all rates except $^{18}$O(n,$\gamma$)$^{19}$O
where the ratio with the rates determined
from the experimental data of Ref.~\cite{mei96b} is shown.}
\label{fig-o}
\end{figure}
\begin{figure}
\caption[fig7]{Comparison of our reaction rates for neutron capture
on C-isotopes with Hauser-Feshbach
calculations obtained with the code SMOKER.}
\label{fig-smc}
\end{figure}
\begin{figure}
\caption[fig8]{Same as Fig.~\ref{fig-smc} for neutron capture of
N-isotopes.}
\label{fig-smn}
\end{figure}
\begin{figure}
\caption[fig9]{Same as Fig.~\ref{fig-smc} for neutron capture of
O-isotopes.}
\label{fig-smo}
\end{figure}
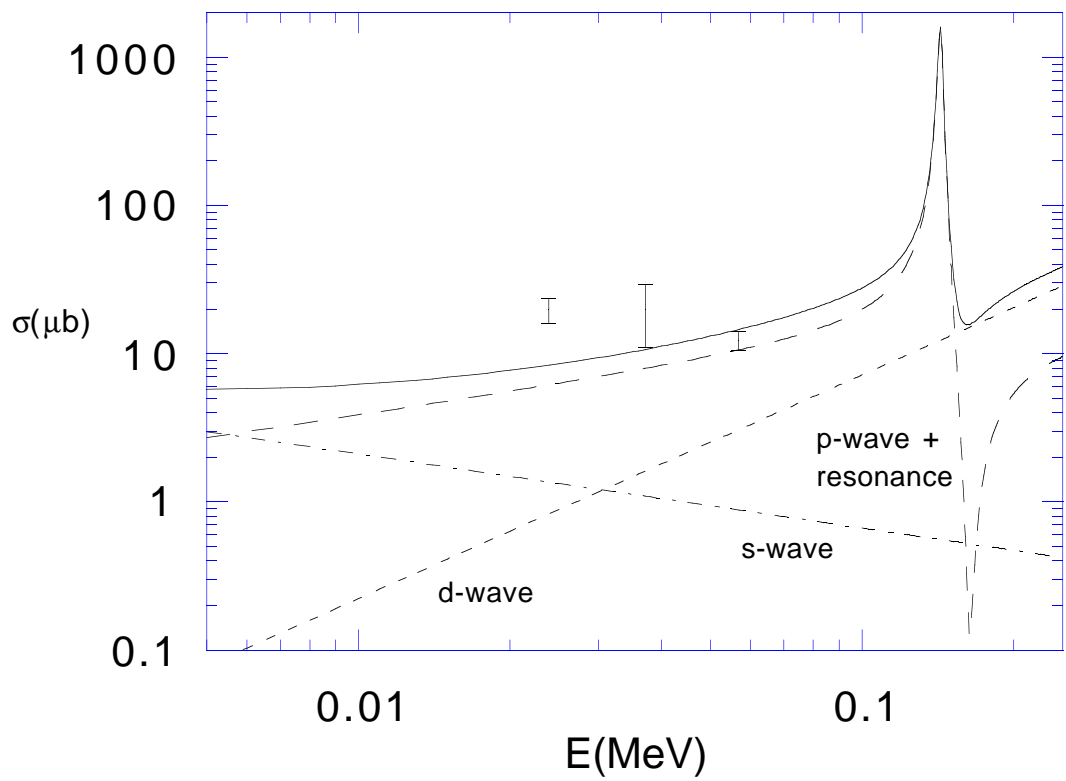
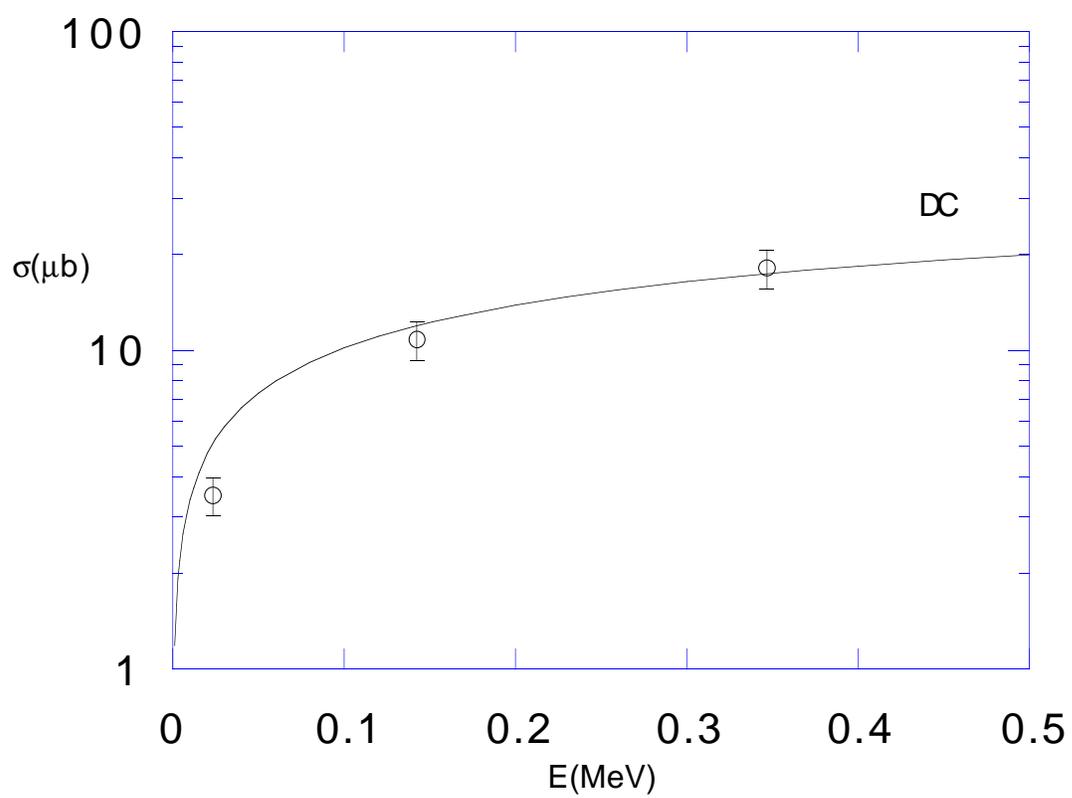
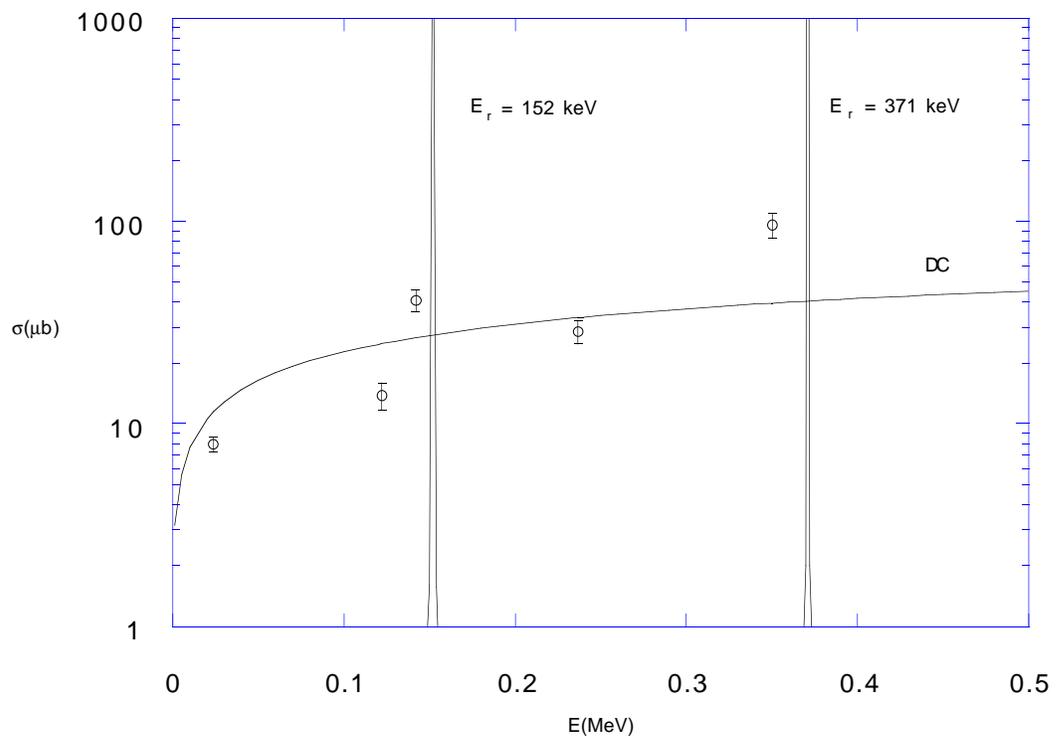
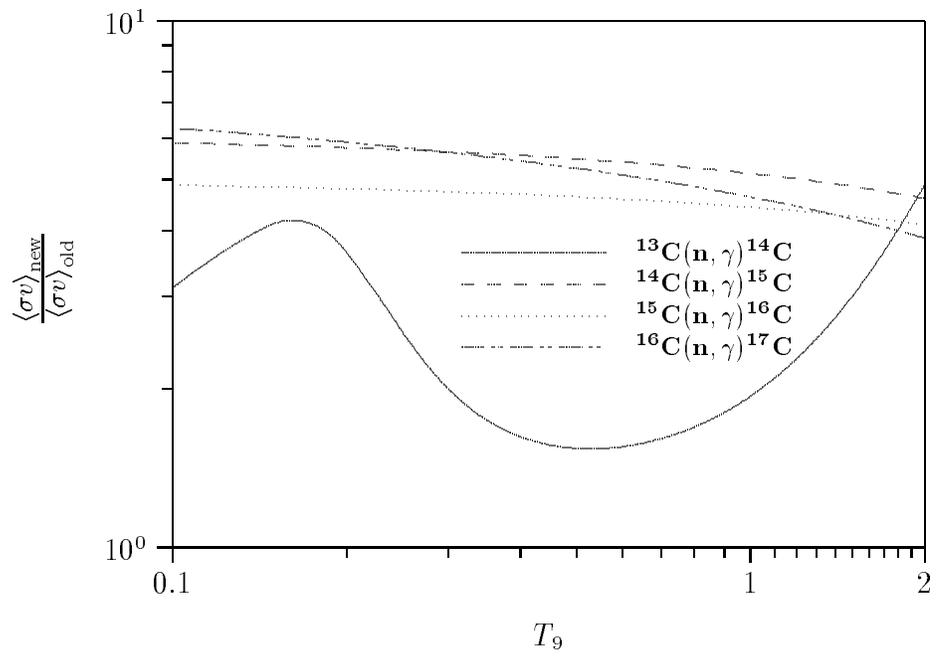
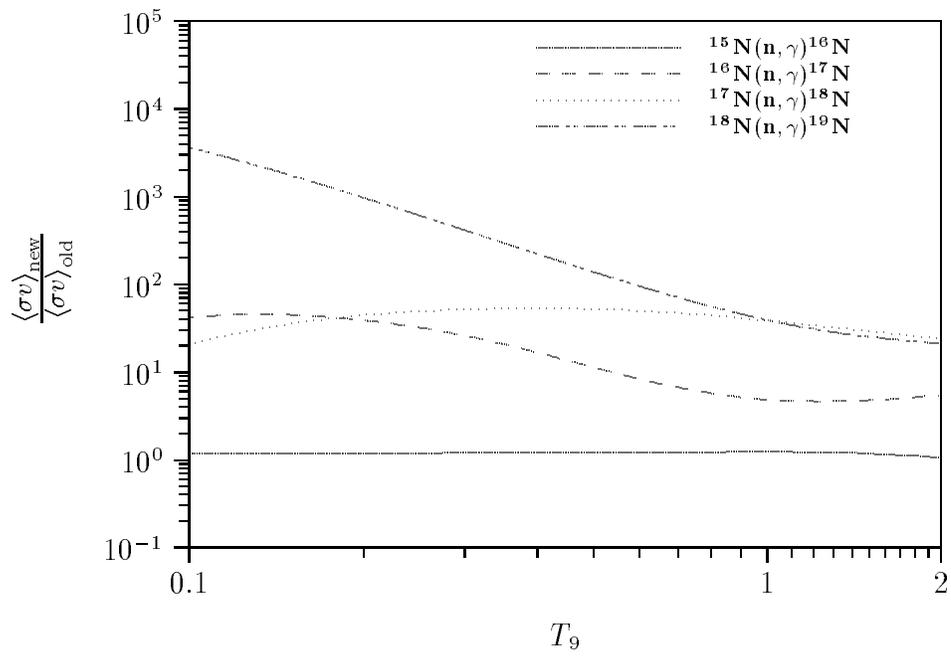
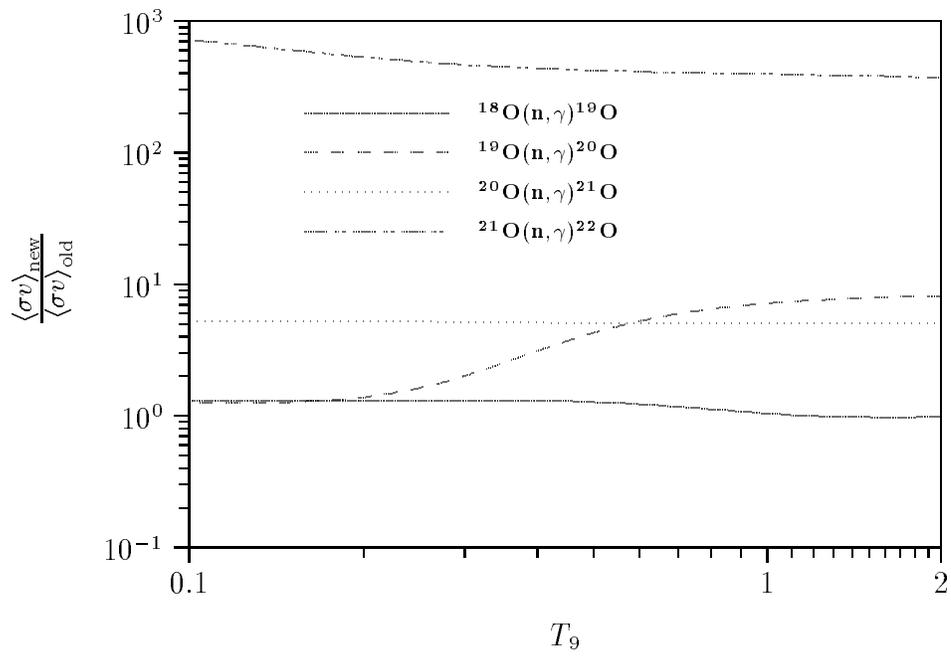
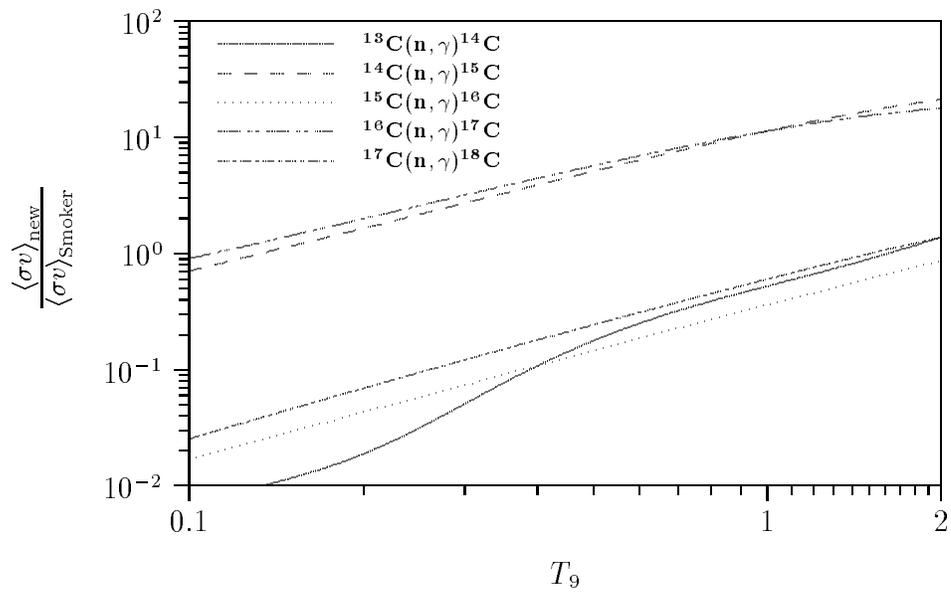
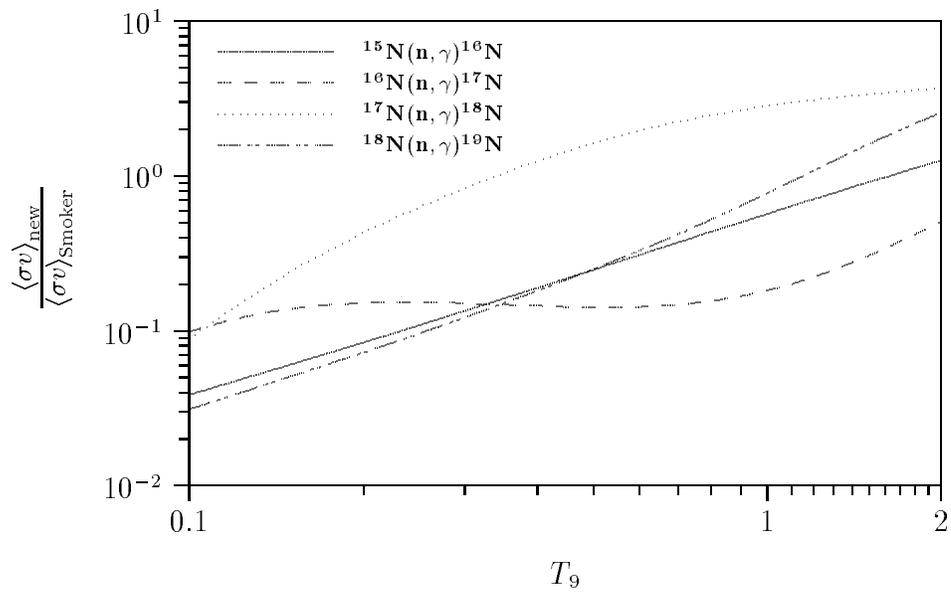
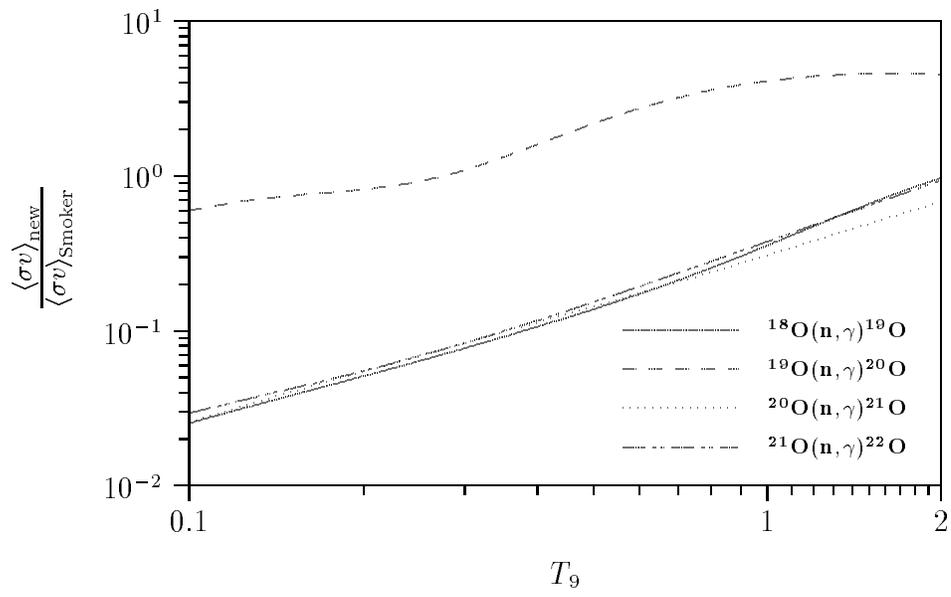
\end{document}